\documentclass[
final            
  ]
  {aipproc}

\layoutstyle{6x9}

\begin{document}

\title{%
{}\footnote{This work was partially supported by Russian Foundation for Basic Research,
projects no. 05-02-17196, 06-02-16659, and 06-02-16353.}
\ QE Neutrino CC Cross Sections off $^{16}$O
}



\author{A. V. Butkevich
}{
  address={Institute for Nuclear Research, 117312 Moscow, Russia}
}

\author{S. A. Kulagin
}{
  address={Institute for Nuclear Research, 117312 Moscow, Russia}
}

\begin{abstract}
The charged-current quasi-elastic scattering of muon neutrino on oxygen
target is computed for neutrino energy between 200 MeV and 2.5 GeV using
different approximations: 
the Plane Wave Impulse Approximation (PWIA),
the Relativistic Distorted Wave Impulse Approximation with relativistic
optical potential (RDWIA), and the Relativistic Fermi Gas Model (RFGM). 
The comparison with RFGM,
which is widely used in data analyses of neutrino experiments, shows that
the RFGM fails completely when applied to exclusive cross section data and
leads to overestimated values of inclusive and total cross sections. 
\end{abstract}

\maketitle


An important source of systematic uncertainties of neutrino oscillation
experiments is related to nuclear effects in neutrino interactions
with detector material. 

The neutrino beams of high intensity cover the energy range from a few hundred
MeV to several GeV. 
Unfortunately, the cross section data in the
relevant energy range are rather scarce and were taken on the targets, which
are not used in neutrino oscillation experiments (i.e., water, iron, lead
or plastic).
In order to simulate the neutrino
detector response in Monte Carlo (MC) codes~\cite{REV1},
a simple nuclear Fermi gas model is commonly used.
This model allows to take into account the effects of the nucleon momentum
distribution, Fermi statistics, and nuclear binding in a simple analytic way.
It should be noted, however, that this model gives an oversimplified picture
of nuclear scattering.
In particular, it lacks to describe the nuclear shell structure,
the final state interaction (FSI) between the outgoing nucleon and residual nucleus, and
the effects of strong short-range nucleon-nucleon (NN) correlations, 
which are known to be important from electron scattering data. 

Motivated by this reasoning
we compute the single-nucleon knockout contribution to the
exclusive cross section, inclusive and total cross sections of
the charged-current QE (anti)neutrino scattering from $^{16}$O in 
PWIA, RDWIA, and the Fermi gas model.
More details about the calculations and the results can be found in Ref.\cite{BK}.
Here we summarize the essential points.  

In the Impulse Approximation (IA), it is assumed that the incoming lepton 
interacts with only one nucleon and 
the nuclear current is approximated by the sum of single-nucleon currents. 
The matrix elements of the nuclear current are treated differently in the considered approximations.

In the PWIA, the 
FSI is neglected and the wave function of the outgoing nucleon is approximated by the plane wave.
In this approximation the nuclear differential cross section is
given in terms of the nuclear spectral function, which
includes the contributions from the nuclear shells as well as the contribution from
the NN-correlations in nuclear ground state.
In our calculations we consider a phenomenological model of the spectral function,
which
incorporates both the single particle nature of the nucleon
spectrum at low energy and high-energy and high-momentum component due to
NN-correlations.
According to electron-nucleus scattering data the 
occupancy of the shell levels of
 $^{16}$O is approximately 75\% on average. 
In this paper we assume that the
missing strength can be attributed to the short-range NN-correlations in the
ground state. 

The RDWIA takes into account the FSI effect, which is described by the wave function of
the outgoing nucleon interacting with the residual nucleus (distorted wave function).
The distorted wave functions are evaluated using a relativized
Schr\"{o}dinger equation for the upper components of Dirac wave functions. 
In numerical calculations
we employ the LEA code~\cite{REV2} adopted for neutrino reactions.
The LEA program was 
initially designed for computing of exclusive proton-nucleus
and  electron-nucleus scattering,
and was successfully tested against $A(e,e'p)$
data (for more detail see \cite{REV2} and references therein).

A complex relativistic optical potential with nonzero imaginary part
generally produces absorption of flux. For the exclusive channel this reflects
the coupling between different open reaction channels. However, for
the inclusive reaction the total flux must conserve. In this work, in order to 
calculate the inclusive and total cross sections,
we use the approach, in which only the real part of the optical potential
is included. 
We also propose a way to estimate the FSI effect on the inclusive cross sections 
in the presence of short-range NN-correlations in the ground state. 

Here we summarize our essential results (for more detail see \cite{BK}):
\begin{itemize}

\item
In RDWIA the reduced exclusive cross sections for $\nu(\bar\nu)$ scattering 
are similar to those of electron scattering (apart from small differences at
low beam energy due to FSI effects for electron and neutrino induced 
reactions).
The calculated electron cross sections are in a good agreement with data.

\item
The inclusive and total cross sections were calculated neglecting the
imaginary part of the relativistic optical potential and taking into
account the effect of NN-correlations in the target ground state
and tested against ${}^{16}$O($e,e'$) scattering data. 
The difference with data is less than 10\% in the peak region.

\item 
The FSI effect reduces the total cross section for about 30\% for
 $E_\nu=200$\ MeV compared to PWIA and decreases with neutrino
energy down to 10\% at 1 GeV.

\item
The effect of NN-correlations further reduces the total cross section for
about 15\% for $E_\nu=200$\ MeV. This effect also
decreases with neutrino energy, down to 8\% at 1 GeV.
\end{itemize}

We also tested the Fermi gas model against $e{}^{16}$O data.
In the RFGM calculation we use the Fermi momentum $p_F=225$~MeV/c and the 
binding energy $\epsilon=27$~MeV. 
The calculation included the Pauli blocking factor.
We found that:

\begin{itemize}
\item 
In the peak region, the RFGM overestimates the measured
inclusive cross section at low momentum transfer 
($|{\bf{q}|}\leq 500$ MeV/c).  
The difference with data is about 20\% at ${|\bf{q}|}=300$ MeV/c and decreases 
as momentum transfer increases. 

\item
The RFGM fails completely when compared to exclusive cross section data.
\item

For the total neutrino cross sections the RFGM result is about
15\% higher than the RDWIA predictions at $E_\nu\sim 1$ GeV.
\end{itemize}

In conclusion, we study QE nuclear interactions for both, neutrino and electron scattering
within different approaches. We found a significant nuclear-model dependence
of the exclusive, inclusive and total cross sections for neutrino energy
$E_{\nu}\leq 1$\ GeV. 
We tested our calculations against the electron data and found that the data favor
the RDWIA results.
This indicates that the application of the RDWIA in Monte Carlo simulations would allow to reduce the systematic
uncertainty in neutrino oscillation parameters.


\begin{theacknowledgments}
We thank J. J. Kelly for useful communications on the LEA code.
S.K. thanks the Organizing Committee of NuInt07 Workshop for 
warm hospitality and local support.
\end{theacknowledgments}

\end{document}